\documentclass[12pt]{article}
\usepackage{graphicx}
\usepackage{cite}

\def \bea{\begin{eqnarray}}
\def \beq{\begin{equation}}
\def \b{{\cal B}}

\def \eea{\end{eqnarray}}
\def \eeq{\end{equation}}

\def \s{\sqrt{2}}
\def \st{\sqrt{3}}
\def \sx{\sqrt{6}}
\def \half{\frac{1}{2}}
\def \3half{\frac{3}{2}}

\def \ok{\overline{K}^0}

\begin{document}
\begin{flushright}
EFI 07-06 \\
TECHNION-PH-2007-05 \\
hep-ph/0702193 \\
February 2007 \\
\end{flushright}
\centerline{\bf ISOSPIN OF NEW PHYSICS IN $|\Delta S|=1$ CHARMLESS $B$ DECAYS}
\bigskip
\centerline{Michael Gronau$^a$ and Jonathan L. Rosner$^b$}
\medskip
\vskip3mm
\centerline{$^a$\it Physics Department, Technion -- Israel Institute of
Technology}
\centerline{\it 32000 Haifa, Israel}
\medskip
\centerline{$^b$\it Enrico Fermi Institute and Department of Physics,
University of Chicago}
\centerline{\it Chicago, Illinois 60637, USA}
\bigskip
\centerline{\bf ABSTRACT}
\medskip
New physics (NP) in charmless strangeness-changing $B$ and $B_s$ decays, which
are dominated by the $b \to s$ penguin amplitudes, can either preserve isospin
or change it by one unit.  A general formalism is presented studying pairs of
processes related to each other by isospin reflection.  We discuss information
on $\Delta I$ in NP amplitudes, provided by time-integrated CP-violating rate
asymmetries in $B^+$ and $B^0$ decays (or in $B_s$ decays), differences between
rates for isospin-reflected processes, and coefficients $S$ of $\sin \Delta m
t$ in time-dependent CP asymmetries.  These four asymmetries in $B^+$ and $B^0$
decays (or five asymmetries in $B_s$ decays) are shown to determine the
magnitude and CP-violating phase of a potential isovector NP amplitude, and the
imaginary part of an isoscalar amplitude, assuming that strong phases in NP
amplitudes are negligible.  This information may be compared with predictions
of specific models, for which we discuss a few examples.
\bigskip

\noindent
PACS Categories:  13.25.Hw, 11.30.Er, 12.15.Ji, 14.40.Nd

\bigskip
\centerline{\bf I.  INTRODUCTION}
\bigskip

The $b \to s$ penguin amplitude is a promising place to look for effects of new
physics in charmless strangeness-changing ($|\Delta S| = 1$) $B$ meson decays
\cite{Gronau:1996rv,Grossman:1996ke,Ciuchini:1997zp,London:1997zk,%
Barbieri:1997kq}.  Observables which can
shed light on its nature include decay rates, direct CP-violating rate
asymmetries, time-dependent CP asymmetries,  differences between
processes related by isospin, and polarization measurements in decays to
two vector mesons.  Using these, one can obtain information on
new physics (NP) entering into the $b \to s$ amplitude. 

The most general $b \to s q \bar q$ operator, where $q$ is a light quark
$u,~d,~s$ or a charmed quark $c$,
is a combination of operators with isospin $\Delta I=0$ and $\Delta
I = 1,~\Delta I_3=0$.  
In the Standard Model (SM), the penguin operator involving a virtual $b \to s$
transition accompanied by a flavor-symmetric $q \bar q$ pair has $\Delta I = 0$
to the extent that it is dominated by intermediate charm or top quarks, while
intermediate on-shell $u$ quarks induce a small $\Delta I = 1$ component.
Transitions treating $b \to s u \bar u$ and $b \to s d \bar d$ differently,
including tree and electroweak penguin amplitudes, contain both $\Delta I =0$
and $\Delta I = 1$ components.  
Four-quark operators associated with NP in
$b \to s q \bar q$ transitions can change isospin by either 0 or 1 unit.  

The present paper is devoted to the question of how one may use all possible
observables in charmless strangeness-changing $B$ meson decays to diagnose the
value of $\Delta I$ in NP operators, regardless of the chiral or color
structure of these operators.  This question has been addressed in the past by
studying a small sample of a few processes including $B\to \phi K$ and $B\to
\pi K$.  Fleischer and Mannel~\cite{Fleischer:2001pc} have shown how to
separate from one another $\Delta I = 0$ and $\Delta I =1$ NP effects
in $B\to \phi K$  by comparing CP asymmetries in charged and neutral $B$
decays.  Sum rules for decay rates and CP asymmetries in $B\to K\pi$ have been
proposed \cite{Gronau:1998ep,Lipkin:1998ie,Atwood:1997iw,Matias:2001ch,%
Gronau:2005gz,Gronau:2005kz,Gronau:2006eb}, whose violation tests NP in
$\Delta I = 1$ transitions.

One of the quantities to be studied together with decay rates and direct CP 
asymmetries is  the coefficient $S$ of the $\sin\Delta m t$ term in
time-dependent CP asymmetries in $B^0$ decays to CP-eigenstates.  A canonical
value, $S= -\eta_{CP}\sin2 \beta$ for final states with CP-eigenvalue
$\eta_{CP}$, is expected approximately for $b \to s$ penguin amplitudes
dominated by top or charm quarks
\cite{Grossman:1996ke,London:1997zk,London:1989ph}.  Small process-dependent
deviations from this value are expected within the SM~\cite{Cheng:2005bg}.
Measurements of this quantity, averaged over a large number of
processes including (quasi) two-body and three-body decays, yield 
$\sin 2 \beta_{\rm eff} = 0.53 \pm 0.05$, to be compared with the value of 
$\sin 2\beta = 0.678 \pm 0.025$ obtained from
$B^0 \to J/\psi K_{S,L}$ \cite{HFAG}.  Thus, the question arises as to what
kind of NP is probed by a deviation of $S$ from its nominal value.
Is it sensitive to new-physics contributions with $\Delta I = 0$?  Does it
reflect {\it only} new $\Delta I = 0$ contributions?  

In this paper we shall address the above questions.  Our purpose is to develop
a general formalism for studying the isospin structure of potential NP effects
in a broad class of (quasi) two-body and multi-body $B$ and $B_s$ decays
mediated by $b\to sq\bar q$ transitions.  We shall see that this information
can be obtained almost completely by combining measurements of $S$ in $B^0$ (or
$B_s$) decays with isospin asymmetry measurements and direct CP asymmetry
measurements in $B^0$ and $B^+$ (or in $B_s$) decays. It will be shown that the
magnitude and CP-violating phase of the $\Delta I=1$ NP term, and the imaginary
part of the $\Delta I=0$ NP term, can be determined, assuming negligible strong
phases in NP amplitudes.  Individual models may be tested against this
information when it becomes available.

In Section II we prove a general relation between charmless $|\Delta S| = 1$
decay amplitudes for pairs of $B$ and $B_s$ decay processes, obtained from each
other under isospin reflection.  This relation permits introducing a general
language suitable for decomposing processes mediated by $b\to sq\bar q$ into
$\Delta I = 0$ and $\Delta I = 1$ contributions.  In Section III we discuss
evidence for $\Delta I=0$ penguin dominance provided by asymmetries measured in
these processes. Section IV presents an expansion of four asymmetries in $B^+$
and $B^0$ decays, and  five asymmetries in $B_s$ decays, in terms of small
ratios involving non-penguin $\Delta I=0$ and $\Delta I=1$ contributions,
showing the amount of information which can be learned from these observables
about NP $\Delta I=0$ and $\Delta I$ amplitudes.  Section V demonstrates the
extraction of NP amplitudes and CP-violating phases in $B\to K\phi, B\to K\pi$
and $B_s\to K\bar K$ decays, assuming that strong phases in these
amplitudes are negligible.  Section VI deals with some aspects of specific
models, including isospin properties of new $b\to sq\bar q$ effective
operators, while Section VII concludes.
\bigskip

\centerline{\bf II.  AMPLITUDES FOR ISOSPIN-REFLECTED DECAYS}
\bigskip

We start by pointing out a general relation between amplitudes for pairs of 
$B$ or $B_s$ decay processes mediated by $\bar b\to \bar sq\bar q$, which are obtained 
from each other under isospin reflection,
\beq
R_I: u \leftrightarrow d~, \bar u \leftrightarrow -\bar d~~.
\eeq
The $\Delta I = 0$ and $\Delta I = 1$ operators, $\bar bs(\bar dd + \bar uu)$
and $\bar bs(\bar dd - \bar uu)$, obtain a relative minus sign under this
transformation.  This suggests that the $\Delta I = 0$ and $\Delta I = 1$
amplitudes behave the opposite way under isospin reflection.  Denoting the
$\Delta I = 0$ and $\Delta I = 1$ contributions in $B^+\to f$ by $B$ 
and $A$, the corresponding $\Delta I = 0$ and $\Delta I = 1$ contributions in
$B^0\to R_If$ are either $B$ and $-A$, or $-B$ and $A$, depending (as we will
show next) on the isospin structure of the final state,
\beq\label{B+-A}
A(B^+\to f) = B + A~~,~~~~A(B^0\to R_If)= \pm(B - A)~~.
\eeq 

Note that while $B$ and $A$ correspond to $\Delta I=0$ and $\Delta I=1$
operators, respectively, these amplitudes do not necessarily involve final
states with a well-defined isospin. The state $f$ in (\ref{B+-A}) may be any
(quasi) two-body or multi-body state.  In our discussion below we will focus
our attention on final states in $B^0$ decays, for which a somewhat low value
of $\sin 2 \beta$ was measured \cite{HFAG}, and on isospin-related charged $B$
decays.  

Our analysis relies largely on Eqs.\ (\ref{B+-A}).  Before proving this general
structure, fixing the signs in the second equation, we will treat several cases
involving specific final states.

The simplest example is the pair of processes  $B^+\to K^+\phi$ and $B^0\to
K^0\phi$, where the final states are pure $I=1/2$, obtaining contributions from
a single $\Delta I=0$ amplitude, $B$, and a single $\Delta I =1$ amplitude, $A$. 
The corresponding Clebsch-Gordan coefficients imply~\cite{Fleischer:2001pc}
\beq
A(B^+\to K^+\phi) = B + A~,~~~~~~A(B^0\to K^0\phi) = B -A~~.
\eeq
Here and below we absorb Clebsch-Gordan coefficients in isospin amplitudes.
Similar decompositions into amplitudes $B\pm A$ apply to such decays as $B \to
K \eta', K\omega$, and $Kf_0(980)$, where the final kaon is accompanied by an
$I=S=0$ meson. 

The frequently-discussed decays $B\to K\pi$, where final states are admixtures
of $I=1/2$ and $I=3/2$, obtain contributions from a single 
$\Delta I=0$ amplitude and two $\Delta I = 1$ amplitudes.
Denoting the final state isospin by a subscript on $B$ and $A$, one 
has~\cite{Lipkin:1991st,Gronau:1991dq}
\bea
-\s A(B^+ \to K^+\pi^0) &=& B_{1/2} + A_{1/2} - 2A_{3/2}~~,\cr
\s A(B^0 \to K^0\pi^0) &=& B_{1/2} - A_{1/2} + 2A_{3/2}~~,\\
A(B^+ \to K^0\pi^+) &=&  B_{1/2} + A_{1/2} + A_{3/2}~~, \cr
-A(B^0 \to K^+\pi^-) &=& B_{1/2} - A_{1/2} - A_{3/2}~~.
\eea
The four processes may thus be grouped into two pairs, with each member of
a pair related to the other by isospin reflection, as in Eq.~(\ref{B+-A}):
\bea\label{Kpi0}
-\s A(B^+ \to K^+\pi^0) &=& B+A~~,~~-\s A(B^0 \to K^0 \pi^0) = -B+A~~,\\
\label{Kpi+-}
A(B^+ \to K^0\pi^+) & = & B+A'~~,~~~~~~~~A(B^0 \to K^+\pi^-) = -B+A'~~.
\eea
Here $B \equiv B_{1/2}$, $A \equiv A_{1/2} - 2 A_{3/2}$, $A' \equiv A_{1/2} +
A_{3/2}$. A similar structure applies to $B\to K\rho$ and $B\to K^*\pi$ decays.
The processes $B^0\to K^0\pi^0$ and $B^0\to K^0\rho^0$ have been used to
obtain the somewhat low averaged value for $\sin 2
\beta_{\rm eff}$~\cite{HFAG}. In all these cases, the coefficient of the
$\Delta I = 0$ amplitude changes sign under isospin reflection, while the
coefficients of the $\Delta I = 1$ amplitudes do not.

Final states in $B^0$ decays involving three neutral kaons are interesting as
they are dominated by $b\to s\bar s s$. Hence in the SM deviations
of $S$ from $-\sin2\beta$ are expected to be very small in
$B^0\to K_SK_SK_S$.  The final state $|f\rangle
\equiv |KK\bar K\rangle$ is a superposition of three isospin states,
$|I_{KK}=0, I_{\rm tot} = \half\rangle,~|I_{KK}=1,~I_{\rm tot}=\half\rangle$
and $|I_{KK}=1,~I_{\rm tot}=\3half\rangle$. Consequently, these decay processes
are described by five independent isospin amplitudes which are functions of the
three kaon momenta.  Suppressing the momentum dependence of the physical
amplitudes and of the five isospin amplitudes,
$A_{\Delta I}^{I_{KK},I_{\rm tot}}$, one finds expressions for the four
physical decay amplitudes~\cite{Gronau:2003ep},
\bea\label{iso1}
A(B^+\to K^+K^+K^-) & = & ~~2A_0^{1,\half} - 2A_1^{1,\half} + 
A_1^{1,\3half}~~,\cr 
\label{iso2}
A(B^0\to K^0K^0\ok) & = & - 2A_0^{1,\half} - 2A_1^{1,\half} + 
A_1^{1,\3half}~~,\\ 
\label{iso3}
A(B^+\to K^+K^0\ok) & = &~~ A_0^{0,\half} - A_0^{1,\half} - 
A_1^{0,\half} + A_1^{1,\half} + A_1^{1,\3half}~~,\cr 
\label{iso6}
A(B^0\to K^0K^+K^-) & = & - A_0^{0,\half} + A_0^{1,\half} - 
A_1^{0,\half} + A_1^{1,\half} + A_1^{1,\3half}~~.
\eea
Amplitudes on the left-hand side correspond to given momenta of the three
particles in the final state. Denoting $B\equiv 2A_0^{1,\half}, A\equiv
- 2A_1^{1,\half} + A_1^{1,\3half}$, $B'\equiv A_0^{0,\half} - A_0^{1,\half},
A' \equiv - A_1^{0,\half} + A_1^{1,\half} + A_1^{1,\3half}$, one has a
situation similar to $B\to K\pi$ for the two pairs consisting of
isospin-reflected processes,
\bea
A(B^+\to K^+K^+K^-) = B + A~,~~& & 
A(B^0\to K^0K^0\ok) =  -B + A~~,\\ 
A(B^+\to K^+K^0\ok) =B' + A'~,~~& &
A(B^0\to K^0K^+K^-) = -B' +A'~~.
\eea

Finally, we consider decays of the form $B\to K\pi\pi$ of which $B^0\to
K_S\pi^0\pi^0$ has been used recently to measure $\sin 2 \beta_{\rm eff}$
\cite{Aubert:2007ub}. Here three pairs of processes are related to one another
by isospin reflection~\cite{Gronau:2005ax}:
\bea
B^+ \to K^+ \pi^+ \pi^- & \Leftrightarrow & B^0 \to K^0 \pi^- \pi^+~~,\\
B^+ \to K^+ \pi^0 \pi^0 & \Leftrightarrow & B^0 \to K^0 \pi^0 \pi^0~~,\\
B^+ \to K^0 \pi^+ \pi^0 & \Leftrightarrow & B^0 \to K^+ \pi^- \pi^0~~.
\eea
The amplitudes of these processes may be expanded in terms of invariant isospin
amplitudes $A_{\Delta I}^{I_{\pi \pi},I_{\rm tot}}$ as follows, where we have
absorbed common factors into the amplitudes as in previous examples:
\bea
A(B^+ \to K^+ \pi^+ \pi^-) &=&  
 A_0^{0,1/2}-A_0^{1,1/2}-A_1^{0,1/2}+ A_1^{1,1/2}+A_1^{1,3/2}- A_1^{2,3/2}
\,,~~\cr A(B^0 \to K^0 \pi^- \pi^+) &=& 
 A_0^{0,1/2}-A_0^{1,1/2}+A_1^{0,1/2}- A_1^{1,1/2}-A_1^{1,3/2}+ A_1^{2,3/2}
\,,~~\\ A(B^+ \to K^+ \pi^0 \pi^0) &=& 
 -A_0^{0,1/2} +A_1^{0,1/2}  -2A_1^{2,3/2}~,~~\cr
A(B^0 \to K^0 \pi^0 \pi^0)&=& 
-A_0^{0,1/2} -A_1^{0,1/2}  +2A_1^{2,3/2}~~,\\
\s A(B^+ \to K^0 \pi^+ \pi^0) &=& 
2A_0^{1,1/2} -2A_1^{1,1/2}+A_1^{1,3/2}+3A_1^{2,3/2}~,~~\cr
\s A(B^0 \to K^+ \pi^- \pi^0)&=& 
2A_0^{1,1/2}+2A_1^{1,1/2}-A_1^{1,3/2}-3A_1^{2,3/2}~~.
\eea 
Defining $B\equiv A_0^{0,1/2}-A_0^{1,1/2}, A\equiv 
-A_1^{0,1/2}+ A_1^{1,1/2}+A_1^{1,3/2}- A_1^{2,3/2}$, $B'\equiv  -A_0^{0,1/2},
A'\equiv A_1^{0,1/2}  -2A_1^{2,3/2}$, $B''\equiv 2A_0^{1,1/2}, A''\equiv 
-2A_1^{1,1/2}+A_1^{1,3/2}+3A_1^{2,3/2}$, the three pairs of physical amplitudes 
can be expressed again as in $B\to K\phi$,
\bea
A(B^+ \to K^+ \pi^+ \pi^-) = B+ A~,~~~~&&
~~~~A(B^0 \to K^0 \pi^- \pi^+) = B - A~,\\
A(B^+ \to K^+ \pi^0 \pi^0) = B' +A'~,~~~&&
~~~~~A(B^0 \to K^0 \pi^0 \pi^0) = B' - A'~,\\
\s A(B^+ \to K^0 \pi^+ \pi^0) = B'' + A''~,~~&&
\s A(B^0 \to K^+ \pi^- \pi^0) = B'' - A''~.~~~
\eea 
As in $B\to K\phi$, under isospin reflection, the coefficients of all $\Delta I
= 1$ amplitudes change sign, while those of the $\Delta I = 0$ amplitudes do
not.

We summarize in Table \ref{tab:sgns} the behavior of the coefficients of the
$\Delta I = 0$ and $\Delta I = 1$ amplitudes for processes $B \to KX$ under
isospin reflection.  We shall now present a general proof for Eq.~(\ref{B+-A}), 
showing that the $\Delta I = 0$ and $\Delta
I = 1$ coefficients always behave in opposite fashion under this reflection.

\begin{table}
\caption{Behavior under isospin reflection of coefficients of invariant
isospin amplitudes describing decays $B \to KX$.
\label{tab:sgns}}
\begin{center}
\begin{tabular}{c c c} \hline \hline
$X$ & $\Delta I = 0$ & $\Delta I = 1$ \\ \hline
$\phi$     & +  & -- \\
$\pi$      & -- & +  \\
$K \bar K$ & -- & +  \\
$\pi \pi$  & +  & -- \\ \hline \hline
\end{tabular}
\end{center}
\end{table}

The proof makes use of the property
\beq
(j_1 j_2 -m_1 -m_2|j_1 j_2 j -m)=(-1)^{j-j_1-j_2} (j_1 j_2 m_1 m_2|j_1 j_2 j m)
\eeq
of Clebsch-Gordan coefficients.
The total isospin $I_{\rm tot} = 1/2$ or 3/2 is formed by the coupling of
$\Delta I = 0,~1$ to $I_B=1/2$, giving rise to a phase $(-1)^{I_{\rm tot} -
\Delta I - 1/2}$ under isospin reflection.  In turn, $I_{\rm tot}$ couples to
the product of $I_X$ and $I_K = 1/2$, giving rise to a phase $(-1)^{I_{\rm tot}
- I_X - 1/2}$.  Finally, there is a phase $\eta_X (-1)^{I_X}$, where $\eta_X$
depends on the specific particles in $X$.  For example, $\eta_X = +1$ for
$X = \phi,~\pi \pi$ and $\eta_X = -1$ for $X=\pi,K \bar K$.  We use the fact
that $2I_{\rm tot} - 1$ is always even, so that all phases
except those due to $\Delta I$ and $\eta_X$ cancel, and the coefficients
of amplitudes describing transitions with $\Delta I = 0,~1$ acquire phases
$\eta_X(-1)^{\Delta I}$ under isospin reflection.

A similar formulation in terms of $\Delta I=0$ and $\Delta I=1$ amplitudes 
$B$ and $A$ applies also to $B_s$ decays, where the initial state has $I=0$ and
final states in $b\to sq\bar q$ transitions are admixtures of $I=0$ and $I=1$. 
Thus, $B_s$ decay amplitudes for pairs of isospin-reflected final states can 
always be expressed in a form similar to Eq.~(\ref{B+-A}). For instance, one has 
\beq\label{BsKKamp}
A(B_s\to K^+K^-)=B+A~~,~~~~~~~~~A(B_s\to K^0\bar K^0)=-B+A~~,
\eeq 
and 
\beq
A(B_s\to K^+\bar K^0\pi^-)=\tilde B+\tilde A~~,~~~~~~~
A(B_s\to K^0K^-\pi^+)=\tilde B-\tilde A~~.
\eeq 

\bigskip

\centerline{\bf III.  TESTS FOR PENGUIN  DOMINANCE}
\bigskip

In the SM, strangeness-changing decays of $B$ and $B_s$ mesons to final 
states consisting of $u, d$ and $s$ quarks are expected to be dominated  by 
$\Delta I=0$ penguin amplitudes involving a CKM-favored factor 
$V^*_{cb}V_{cs}$. NP could, in principle, change this behavior by
introducing comparable $\Delta I=0$ or $\Delta I=1$ contributions involving 
new CP-violating phases. In this section we discuss circumstantial evidence
showing that this is not the case.  Namely, potential $\Delta I=0$ and $\Delta
I=1$ contributions beyond the Standard Model are 
most likely much smaller than the Standard Model penguin amplitudes. 

Consider a pair of $B^+$ and $B^0$ decay processes as discussed in Section II, 
$B^+\to KX$ and $B^0\to K_{R_I}X_{R_I}$, related to each other under isospin 
reflection. Here $K$ may be a $K^+$ or a $K^0$, and $X$ is an arbitrary
charmless and nonstrange hadronic state with a corresponding total charge.  As
we have shown, the amplitudes for this pair of processes can be expressed in a
model-independent way as 
\beq
A(B^+\to KX) = B + A~~,~~~~~~~A(B^0\to K_{R_I}X_{R_I}) = \pm(B-A)~~,
\eeq
where {\em $B$ and $A$ are $\Delta I=0$ and $\Delta I =1$ amplitudes depending
on $X$}.  The corresponding decay rates are 
\beq
\Gamma_+ \equiv \Gamma(B^+\to KX) = |B+A|^2~,~~~~~~~
\Gamma_0 \equiv \Gamma(B^0\to K_{R_I}X_{R_I})= |B-A|^2~~,
\eeq
where inessential kinematic factors have been omitted including phase space
integration in the case of three-body or other multi-body decays.  The rates
for $B^-$ and $\bar B^0$ decays into charge-conjugate final states are
\beq
\Gamma_- \equiv \Gamma(B^-\to\bar K\bar X)= |\bar B + \bar A|^2~,~~~~~~
\Gamma_{\bar 0} \equiv \Gamma(\bar B^0\to\bar K_{R_I}\bar X_{R_I})
 = |\bar B - \bar A|^2~~.
\eeq
The charge-conjugated amplitudes $\bar B$ and $\bar A$ are related to $B$ and
$A$ by a change in sign of all weak phases, whereas strong phases are left
unchanged.

In the Standard Model, the isoscalar amplitude $B$ contains a dominant penguin 
contribution, $B_P$, with a CKM factor  $V^*_{cb}V_{cs}$.
The residual isoscalar amplitude,
\beq
\Delta B\equiv B-B_P~~,
\eeq
and the amplitude $A$, consist each of smaller contributions.  These include
terms with a much smaller CKM factor $V^*_{ub}V_{us}$, and a higher order
electroweak penguin amplitude with CKM factor $V^*_{tb}V_{ts}$.  In $B\to K\pi$
decays, for instance, the largest term of the first kind is a color-favored
tree amplitude. This amplitude and a  comparable electroweak penguin term are 
suppressed by about an order of magnitude relative to $B_P$
\cite{Gronau:1994rj,Ciuchini:1997rj,Keum:2000wi,Beneke:2001ev,Bauer:2005kd}.
In $B\to K\phi$, the first amplitude is even smaller. Thus, in general one
expects 
\beq\label{hierarchy}
|\Delta B|  \ll  |B_P|~~,~~~~|A|\ll |B_P|~~.
\eeq 

\begin{table}
\caption{Isospin-dependent asymmetries $A_I$
defined in (\ref{A_I}),
CP-asymmetries $A^{+,0}_{CP}$
defined in (\ref{CP-asym}), and mixing-induced CP asymmetries $S$ for $B\to
KX$~\cite{HFAG}.
\label{tab:asym}}
\begin{center}
\begin{tabular}{c c c c c} \hline \hline
$X$ & $A_I$ &  $A^+_{CP}$ & $A^0_{CP}$ & $-\eta_{CP}S$\\ \hline
$\phi$  &  $-0.037 \pm 0.077$  & $0.034 \pm 0.044$
 & $-0.01\pm 0.13$ &$0.39\pm 0.18$\\
$\eta'$  & $-0.001 \pm 0.033$ & $0.031 \pm 0.026$
 & $0.09 \pm 0.06$ & $0.61\pm 0.07$\\
$\omega$ & $0.14 \pm 0.07$ & $0.05 \pm 0.06$
 & $0.21\pm 0.19$ & $0.48\pm 0.24$ \\
$f_0(980)$  & $0.18 \pm 0.08$  & $-0.026^{+0.068}_{-0.064}$
 & $0.02\pm 0.13$ & $0.42\pm 0.17$\\
$\pi^0$  & $0.087 \pm 0.038$ & $0.047 \pm 0.026$
 & $-0.12\pm 0.11$ & $0.33\pm 0.21$\\
$\pi^+$  & $0.051 \pm 0.026$ & $0.009 \pm 0.025$
 & $-0.097\pm 0.012$ & --\\
$\rho^0$ & $-0.16 \pm 0.10$ & $0.31^{+0.11}_{-0.10}$
 & $-0.64\pm 0.46$ & $0.20\pm 0.57$ \\
 $\rho^+$ & $-0.14\pm 0.12~^a$ & $-0.12\pm 0.17~^a$ & $0.17^{+0.15}_{-0.16}$ & --\\
$K^+K^-$ & $0.25 \pm 0.07~^b$ & $-0.02\pm 0.04$
 & $-0.15\pm 0.09$ &  $0.58^{+0.18}_{-0.13}$\\
$K^0\bar K^0$ & $0.018\pm 0.080~^b$ & $-0.04 \pm 0.11$
 & $0.14\pm 0.15~^c$ & $0.58\pm 0.20~^c$ \\
$\pi^+\pi^-$  & $0.064\pm 0.039$ & $0.023 \pm 0.025$ & -- & -- \\ 
$\pi^0\pi^0$ & -- & -- & $-0.23 \pm 0.54~^d$ & $-0.72 \pm 0.71~^d$\\ 
\hline \hline
\end{tabular}
\end{center}
\leftline{$^a$ Values of branching ratio and asymmetry for $B^+\to K^0\rho^+$ 
taken from Ref.~\cite{Aubert:2007mb}.}
\leftline{$^b$ Values obtained by subtracting $B\to K\phi$ contributions, 
assuming that the sub-} 
\leftline{tracted amplitudes are symmetric with respect to interchanging
$K^+$ and $K^-$ mo-} 
\leftline{menta and $K^0$ and $\bar K^0$ momenta~\cite{Gronau:2005ax}.}
\leftline{$^c$ Values include measurements reported in Ref.~\cite{Aubert:2007me}.}
\leftline{$^d$ Values from Ref.~\cite{Aubert:2007ub}.} 
\end{table}

Tests for the hierarchy (\ref{hierarchy}) are provided by an isospin-dependent 
asymmetry,  
\beq\label{A_I}
A_I \equiv \frac{\Gamma_+ + \Gamma_- - \Gamma_0 - \Gamma_{\bar 0}}
 {\Gamma_+ + \Gamma_- + \Gamma_0 + \Gamma_{\bar 0}}~~,
 \eeq
and by two CP-violating asymmetries, in charged and neutral $B$ decays, 
\beq\label{CP-asym}
A^+_{CP}\equiv\frac{\Gamma_- - \Gamma_+}{\Gamma_-+\Gamma_+}~~,~~~~~
A^0_{CP}\equiv\frac{\Gamma_{\bar 0}-\Gamma_0}{\Gamma_{\bar 0} + \Gamma_0}~~.
\eeq
The asymmetries $A_I$ and $A^{+,0}_{CP}$ are expected to be small in the CKM
framework, of order $2|A|/|B|$ and $2|\Delta B|/|B_P|$.  In contrast,
potentially large contributions to $\Delta B$ and $A$ from NP,
comparable to $B_P$, would most likely lead to large asymmetries of order one.
An unlikely exception is the case when both $\Delta B/B_P$ and $A/B_P$ are
purely imaginary, or almost purely imaginary.   This would require very 
special circumstances such as fine-tuning in specific models.  

Current values for the asymmetries $A_I$, $A^+_{CP}$, $A^0_{CP}$, and the 
mixing-induced asymmetry $S$ multiplied  by minus the CP eigenvalue
$-\eta_{CP}$ of the final state in corresponding $B^0$ decays are quoted in
Table \ref{tab:asym}~\cite{HFAG}. We have used the ratio of $B^+$ and $B^0$ 
lifetimes, $\tau_+/\tau_0=1.076 \pm 0.008$, for translating ratios of $B^+$ and
$B^0$ branching ratios into ratios of corresponding decay rates.  The
asymmetries $A_I$ for $X=K^+K^-$ and $X= K^0\bar K^0$ were obtained after
subtracting $B\to K\phi$ contributions in $B^+$ and $B^0$ decays, assuming that
the subtracted amplitudes are symmetric with respect to interchanging $K^+$ and
$K^-$ momenta and $K^0$ and $\bar K^0$ momenta.  This symmetry assumption,
which is required for obtaining $\b(B^+\to K^+K^0\bar K^0)$ from the measured
$\b(B^+\to K^+K_SK_S)$ and $\b(B^0\to K^0K^0\bar K^0)$ from 
$\b(B^0\to K_SK_SK_S)$~\cite{Gronau:2005ax}, is motivated by data, but may hold
only approximately~\cite{Garmash:2004wa,Aubert:2006nu}. This may explain the 
somewhat large  value of $A_I$ for $X=K^+K^-$.  

In general, CP asymmetries in charged $B$ decays involve smaller experimental 
errors than corresponding asymmetries in neutral $B$ decays, which require
flavor tagging and time-dependent measurements.  An exception is the
self-tagged mode $B^0\to K^+\pi^-$, where a nonzero asymmetry has been measured
with a small error.  We will focus attention on the two asymmetries $A^+_{CP}$
and $A_I$  which involve the smallest errors.

Central values measured for the asymmetries $A^+_{CP}$ are in most cases  at a
level of several percent with comparable errors, implying asymmetries of no
more than ten or fifteen percent. Larger asymmetries, at a level of twenty or
thirty percent, are possible in $B^+\to \rho^0K^+, \rho^+K^0, K^+K^0\bar K^0$.
A similar situation occurs in isospin asymmetries $A_I$, which are fairly small
for several final states involving $X=\phi, \eta', \pi^0, \pi^+, K^0\bar K^0,$
$\pi^+\pi^-$, while larger asymmetries at a level of twenty or thirty percent
are possible for other final states. (Note that the asymmetries $A_I$ for
$X=\omega, f_0, \rho^0$ depend on modeling the Dalitz plot for $B^{+,0}\to
K^{+,0}\pi^+\pi^-$~\cite{Abe:2005qe,Aubert:2006qu,Aubert:2005ce}.) Thus, since
$A^+_{CP}$ and $A_I$ are expected to be of order $2|\Delta B|/|B_P|$ and
$2|A|/|B|$, this confirms the hierarchy (\ref{hierarchy}),
excluding NP contributions to $\Delta B$ and $A$ comparable to $B_P$. 

A test for the smallness of the magnitude of $A/B$ regardless of its phase is
provided by a sum rule suggested several years ago for the four $B\to K\pi$
decay rates~\cite{Gronau:1998ep,Lipkin:1998ie}. Using Eqs.~(\ref{Kpi0}) and
(\ref{Kpi+-}), one has
\bea
2\Gamma(K^+\pi^0) + 2\Gamma(K^0\pi^0) & = & 2(|B|^2+|A|^2)~~,\cr
\Gamma(K^0\pi^+) + \Gamma(K^+\pi^-) & = & 2(|B|^2 +|A'|^2)~~.
\eea
This implies a sum rule also for charge-averaged decay rates, $\bar\Gamma(B\to
f)\equiv [\Gamma(B\to f) +\Gamma(\bar B\to \bar f)]/2$,
\beq
2\bar\Gamma(K^+\pi^0) + 2\bar\Gamma(K^0\pi^0) =
\bar\Gamma(K^0\pi^+) + \bar\Gamma(K^+\pi^-)~~.
\eeq
A quadratic multiplicative correction to the sum rule, $(|A'|^2-|A|^2)/|B|^2$,
is  a few percent in the Standard Model
\cite{Bauer:2005kd,Gronau:2003kj,Beneke:2003zv}.  Since $A$ and $A'$ 
involve two independent $\Delta I=1$
amplitudes, the fact that this sum rule holds experimentally within
$5\%$~\cite{Gronau:2006xu} seems to rule out large $\Delta I=1$ NP
contributions of order $B_P$, which would have to cancel within a few percent.
\bigskip

\centerline{\bf IV.  LINEAR EXPANSION IN $\Delta B$ and $A$}
\bigskip

As we have shown, the hierarchy (\ref{hierarchy}) expected in the SM is 
confirmed by experiments, and can therefore be assumed to hold also
in the presence of NP. Thus, we will expand the four asymmetries in
Table \ref{tab:asym} to leading order in $\Delta B/B_P$ or $A/B_P$.  We will
take by convention the dominant penguin amplitude $B_P$ to have a zero weak
phase and a zero strong phase, referring all other strong phases to it. 
Writing
\beq\label{Bconvention}
B = B_P + \Delta B~~,~~~\bar B = B_P + \Delta \bar B~~,
\eeq
and defining for $B^0$ decays to CP-eigenstates
\beq \label{eqn:S}
S = \frac{2 {\rm Im} \lambda}{1 + |\lambda|^2}~~,~~~\lambda \equiv \eta_{CP}
\frac{\bar B - \bar A}{B - A} e^{- 2 i \beta}~~~,
\eeq
we find 
\bea \label{eqn:obs1}
\Delta S & \equiv & -\eta_{CP}S - \sin 2 \beta = \cos 2 \beta\left [\frac{{\rm Im}(\bar A - A)}
{B_P} - \frac{{\rm Im}(\Delta \bar B - \Delta B)}{B_P}\right ]~~,\\
\label{eqn:obs2}
A_I   &=& \frac{{\rm Re}(\bar A + A)}{B_P}~~,\\
\label{eqn:obs3}
A^+_{CP} &=& \frac{{\rm Re}(\bar A- A)}{B_P} + \frac{{\rm Re}(\Delta \bar B
 - \Delta B)}{B_P}~~,\\
\label{eqn:obs4}
A^0_{CP} &=& -\frac{{\rm Re}(\bar A- A)}{B_P} + \frac{{\rm Re}(\Delta \bar B
 - \Delta B)}{B_P}~~.
\eea

Eq.~(\ref{eqn:obs1}) is familiar from its implication in the SM, where one
considers the effect on $S$ of a small amplitude with weak phase $\gamma$ and a
strong phase $\delta$ relative to the dominant penguin amplitude $B_P$, $\Delta
B - A = |\Delta B-A| e^{i\delta}e^{i\gamma}$. When inserted into
(\ref{eqn:obs1}) this implies
\beq
\Delta S = 2\frac{|\Delta B-A|}{B_P}\cos 2\beta\sin\gamma\cos\delta~~,
\eeq
which is a well-known result~\cite{Gronau:1989ia}.
This result is used to calculate deviations of $-\eta_{CP}S$ from 
$\sin 2\beta$, and to argue that $\Delta S>0$ when one 
expects $|\delta|< \pi/2$~\cite{Cheng:2005bg}.

The relations (\ref{eqn:obs1})--(\ref{eqn:obs4}) tell us several things:

(1) The $\Delta I = 0$ and $\Delta I = 1$ contributions in $A_{CP}$ may be
separated from one another by taking sums and differences
\cite{Fleischer:2001pc}:
\bea\label{ACP0}
A^{\Delta I=0}_{CP} & \equiv & \frac{1}{2}(A^+_{CP} + A^0_{CP}) = 
\frac{{\rm Re}(\Delta \bar B - \Delta B)}{B_P}~~,\\
\label{ACP1}
A^{\Delta I=1}_{CP} & \equiv & \frac{1}{2}(A^+_{CP} - A^0_{CP})  =
\frac{{\rm Re}(\bar A - A)}{B_P}~~.
\eea

(2) One can separate the $\Delta I = 1$ terms ${\rm Re}A/B_P$ and ${\rm Re}
\bar A/B_P$ from one another using information from $A^{\Delta I=1}_{CP}$
and $A_I$.

(3) The deviation of $S$ from its nominal value of $- \eta_{CP} \sin(2\beta)$
is governed by an {\it imaginary} part of a combination of $\Delta I = 0$ and
$\Delta I = 1$ terms.  Thus, without the help of a specific model, it is
impossible to determine whether any deviation of $S$ from its nominal value is
due to $\Delta I = 0$ or $\Delta I = 1$ or a combination. One may, however,
test the predictions of specific models for the four asymmetries
(\ref{eqn:obs1})--(\ref{eqn:obs4}).

A similar expansion in terms of small ratios of amplitudes can be performed for 
asymmetries in $B_s$ decays.
Consider, for instance, the pair of isospin-reflected decays $B_s\to K^+K^-$ and 
$B_s\to K^0\bar K^0$, whose amplitudes are given in Eq.\,(\ref{BsKKamp}),
where the observed final  state in the second process  is $K_SK_S$.
In this case, one may measure in principle {\em five} asymmetries (instead of four in $B^+$
and $B^0$ decays): an isospin asymmetry, a pair of direct asymmetries $A_{CP}$ in decays 
to charged and neutral kaons, and a pair of asymmetries $S$ in these decays. 
The first three asymmetries are given by expressions as in Eqs.~(\ref{eqn:obs2})--(\ref{eqn:obs4}).
Denoting the small  phase of $B_s$-$\bar B_s$ mixing by $2\chi$~\cite{Aleksan:1994if}, 
\beq
\chi \equiv {\rm Arg}\left(-\frac{V_{cb}V^*_{cs}}{V_{tb}V^*_{ts}}\right )~~,
\eeq
where in the SM~\cite{Silva:1996ih}
\beq
\sin\chi = \left |\frac{V_{us}}{V_{ud}}\right |^2\frac{\sin\beta\sin\gamma}{\sin(\beta + \gamma)}~~,
\eeq
the other two asymmetries are given by
\bea\label{Sk+k-}
S_{K^+K^-} - \sin 2\chi & = & \cos 2\chi \left [\frac{{\rm Im}(\bar A - A)}
{B_P} + \frac{{\rm Im}(\Delta \bar B - \Delta B)}{B_P}\right ]~~,\\
\label{Sk0k0}
S_{K_SK_S} - \sin 2\chi & = & \cos 2\chi \left [-\frac{{\rm Im}(\bar A - A)}
{B_P} + \frac{{\rm Im}(\Delta \bar B - \Delta B)}{B_P}\right ]~~.
\eea
Thus, a fourth lesson is the following:

(4) In contrast to $B^0$ and $B^+$ decays, these two deviations of $S$ 
from $\sin 2\chi$ enable separating imaginary parts of contributions from 
$\Delta I=0$ and $\Delta I=1$ amplitudes. The difficult part is measuring 
time-dependence in $B_s\to K_SK_S$.
Alternatively, one may measure time-dependent CP asymmetries $S$ in 
$B_s\to K^{*+}K^{*-}$ and $B_s\to K^{*0}\bar K^{*0}$, where the decay time
is obtained by reconstructing $K^*\to K\pi$. Separating between CP-even and 
CP-odd final states requires analyzing the distributions of transversity angles in
$K^*$ decays.

\bigskip

\centerline{\bf V.  EXTRACTING NEW PHYSICS AMPLITUDES}
\bigskip

Under some circumstances one can carry the above discussion further without
referring to a specific model, by assuming that the strong phases associated
with NP amplitudes are small relative to those of the SM and can be
neglected~\cite{Datta:2004re,Baek:2004rp,Datta:2004jm}.  This assumption is
reasonable since rescattering from a leading $b\to sc\bar c$ amplitude is
likely the main source of strong phases, while rescattering from a smaller
$b\to sq\bar q$ NP amplitude is then a second-order effect.  Under this
assumption, all magnitudes and weak phases of NP contributions in $A$ and
$\Delta B$ can be combined into one effective magnitude and weak phase which
changes sign under CP conjugation.  We note that while this assumption is
intuitive and plausible, it is an approximation which must be confronted by
data. (See the discussion at the end of Section V.B.)

The amount of SM contributions in $A$ and $\Delta B$, involving a CKM phase 
different than in $B_P$, is process-dependent. These contributions are very
small in $B\to K\phi$, implying a tiny value for $\Delta S$ within the SM.
They are larger in $B\to K\pi$, leading on the one hand to a potentially larger
value of $\Delta S$ in the SM, but on the other hand to a more subtle treatment
of NP effects in these processes. We will now demonstrate the extraction of NP
amplitudes in these two cases, where the current measurement of $\Delta S$ is
$1.6\sigma$ away from zero for each process (see Table \ref{tab:asym}),
and in $B_s\to K\bar K$. 
\bigskip

\leftline{\bf A.  $B \to K \phi$}
\bigskip

Consider first the decays $B^+ \to K^+ \phi$ and $B^0 \to K^0 \phi$, where
both amplitudes are dominated by the same penguin term $B_P$, with weak phase
$\simeq {\rm Arg}(V_{cb}V^*_{cs}) \simeq 0$. 
[There will be a negligible contamination in $B_P$ of order $2\%$ (the ratio of 
CKM elements) from $V_{ub}V^*_{us}$.] The SM terms in 
$\Delta B$ and $A$ include subleading electroweak penguin (EWP) 
contributions~\cite{Dighe:1997wj}, with a vanishing weak phase similar to
$B_P$. (We neglect a small annihilation amplitude in $B\to K^+\phi$.) The SM
$\Delta I=0$ EWP amplitude, which can acquire a strong phase by rescattering
from $B_P$, will be included in the definition of $B_P$. In contrast, the SM
$\Delta I=1$ EWP term can be assumed to get no strong phase, since it involves
no rescattering from the dominant $\Delta I=0$ amplitude $B_P$. Thus, in our
definition the amplitude $\Delta B$ is purely NP, while $A$ involves a SM EWP
amplitude and NP contributions, both of which are assumed to involve no strong
phases. 

We will now study the amount of information which can be learned about
magnitudes and weak phases of isoscalar and isovector NP amplitudes, using the
four observables (\ref{eqn:obs1}) (\ref{eqn:obs2}) (\ref{ACP0}) and
(\ref{ACP1}). Since these four asymmetries are first order in small ratios of
amplitude, we can take $B_P$ as given by the square root
of $\Gamma(B\to K\phi)$, thereby neglecting second order terms.
We will assume that the EWP contribution to the $\Delta I=1$ 
amplitude $A$ is calculable within the SM~\cite{Beneke:2003zv,Chen:2001pr},
or can be obtained independently by fitting within flavor SU(3) other $B$ decay 
rates and asymmetries~\cite{Chiang:2003pm}.

In our convention (\ref{Bconvention}), where the strong phase of $B_P$ is set 
equal to zero,  $\Delta B$ and $A$ have the same strong phase $\delta$,
and involve weak phases $\phi_B$ and $\phi_A$, respectively, 
\beq\label{DeltaB,A}
\Delta B = |\Delta B|e^{i\delta}e^{i\phi_B}~~,~~~~~
A = |A|e^{i\delta}e^{i\phi_A}~~.
\eeq
In the charge-conjugated amplitudes $\Delta\bar B$ and $\bar A$ the phase
$\delta$ is left unchanged, while $\phi_B$ and $\phi_A$ change signs. 
This leads to
\beq\label{r}
\frac{{\rm Re}(\bar A - A)}{{\rm Re}(\Delta\bar B- \Delta B)} =
\frac{{\rm Im}(\bar A - A)}{{\rm Im}(\Delta\bar B- \Delta B)} =
\frac{|A|\sin\phi_A}{|\Delta B|\sin\phi_B}
\equiv r~~,
\eeq
and
\bea\label{eqn:ReB/ImB}
\frac{{\rm Re}(\Delta \bar B - \Delta B)}{{\rm Im}(\Delta \bar B - \Delta B)}
& = & - \tan \delta~~,\\
\label{eqn:imB}
{\rm Im}(\Delta \bar B - \Delta B)
 & = & - 2 |\Delta B| \sin \phi_B \cos \delta~~,\\
 \label{eqn:ReA}
{\rm Re}(\bar A + A) & = & 2|A|\cos\phi_A\cos\delta~~.
\eea

The parameter $r$ is seen to be determined by the ratio of $\Delta I=1$ and $\Delta I=0$ 
CP asymmetries given in Eqs.~(\ref{ACP1}) and (\ref{ACP0}),
\beq
r = \frac{A^{\Delta I=1}_{CP}}{A^{\Delta I=0}_{CP}}~~.
\eeq
The second ratio in (\ref{r}) implies
\beq
\Delta S = \frac{{\rm Im}(\Delta \bar B - \Delta B)}{B_P}(r-1)\cos 2 \beta~~,
\eeq
leading by (\ref{eqn:ReB/ImB}) to a determination of $\tan\delta$,
\beq
\tan\delta = \frac{A^{\Delta I=0}_{CP}}{\Delta S}(1-r)\cos 2\beta~~.
\eeq
This fixes $\tan\phi_A$ from the ratio of the other two asymmetries,
\beq
\tan\phi_A\tan\delta = \frac{A^{\Delta I=1}_{CP}}{A_I}~~.
\eeq 
Once $\delta$ and $\phi_A$ are given (mod $\pi$), knowledge of $B_P$ and a
measurement of $A_I$ or $A^{\Delta I=1}_{CP}$ gives $|A|$ through
(\ref{eqn:obs2}) and (\ref{eqn:ReA}) or (\ref{ACP1}) and (\ref{r}). (Note that
$A$ includes a potential NP contribution and a SM EWP amplitude, both of which
were assumed to have equal strong phases.) Similarly, a measurement 
of $A^{\Delta I=0}_{CP}$ and $\Delta S$ permit one to obtain
${\rm Re}(\Delta\bar B-\Delta B)$ and ${\rm Im}(\Delta \bar B - \Delta B)$, 
each of which yields $|\Delta B| \sin \phi_B$ from Eqs.\ (\ref{eqn:ReB/ImB})
and (\ref{eqn:imB}). The combination $|\Delta B| \cos \phi_B$ adds 
coherently to $B_P$ and cannot be fixed independently.
\bigskip

\leftline{\bf B.  $B \to K \pi$} 
\bigskip

In $B\to K\pi$, an isolation of NP  contributions in $\Delta B$ and $A$
requires taking into account two subleading SM amplitudes involving a weak
phase ${\rm arg}(V^*_{ub}V_{us})=\gamma$~\cite{Gronau:1994rj}. These are a
color-favored tree amplitude $T$ contributing to $B^0 \to K^+ \pi^-$ and $B^+
\to K^+ \pi^0$, and a color-suppressed tree amplitude $C$ contributing to $B^+
\to K^+ \pi^0$ and $B^0\to K^0 \pi^0$. We consider the two pairs of
isospin-reflected processes $(B^+\to K^+\pi^0, B^0\to K^0\pi^0)$ and $(B^+\to
K^0\pi^+, B^0\to K^+\pi^-)$.  The four measured asymmetries in the first pair
and the three asymmetries in the second pair (see Table~\ref{tab:asym}) combine
an interference of the dominant penguin amplitude $B_P$ with $T$ and $C$ and
with potential NP contributions.

Using the notations of Eqs.~(\ref{Kpi0}) and (\ref{Kpi+-}) and expressing NP
contributions explicitly, one has~\cite{Gronau:1994rj,Gronau:2003kj}
\bea\label{DB-Kpi}
\Delta B & = & \half T + \Delta B_{NP}~~,\\
\label{A-Kpi}
A & = & \half T + C + P_{EW} + \half P^c_{EW} + A_{NP}~~,\\
A' & = & -\half T - \half P^c_{EW} + A'_{NP}~~,
\eea
where $P_{EW}$ and $P^c_{EW}$ are color-favored and color-suppressed 
EWP amplitudes.
The dominant amplitude $B_P$ with weak phase $\simeq {\rm Arg}(V_{cb}V^*_{cs}) 
\simeq 0$ includes a small EWP contribution, $\frac{1}{6}P^c_{EW}$. An  
annihilation amplitude with weak phase $\gamma$, expected to be 
smaller than $T$ and $C$~\cite{Gronau:1994rj,Bauer:2005kd}, is included in 
the definition of $T$ in $B$ and $A$, but is neglected in $A'$. 

Proceeding as before under the assumption that subleading NP amplitudes involve 
negligible strong phases, we apply the same assumption to all subleading 
contributions including $T, C, P_{EW}$ and $P^c_{EW}$, which involve no 
rescattering from the dominant amplitude $B_P$. In our convention, where the 
strong phase of $B_P$ vanishes, expression (\ref{DeltaB,A}) for $\Delta B$ and
$A$ hold also in $B^+\to K^+\pi^0$ and $B^0\to K^0\pi^0$. Consequently, the
entire analysis following these expressions holds too, where $B_P$ is now given
by the square root of $\Gamma(B^+\to K^0\pi^+)$ (which involves tiny second 
order corrections). Thus, measurements of the four asymmetries, $A_I, A^+_{CP},
A^0_{CP}$ and $\Delta S$, in the pair $(B^+\to K^+\pi^0, B^0\to K^0\pi^0)$
leads to a determination of the magnitude $|A|$ and weak phase $\phi_A$ of the
amplitude $A$ and of the quantity $|\Delta B|\sin\phi_B$.  Assuming, as in the
case of $B\to K\phi$, that the magnitudes of all subleading SM amplitudes are
calculable~\cite{Keum:2000wi,Beneke:2001ev,Bauer:2005kd} (or can be fitted
independently within flavor SU(3) to other decays into two 
pseudoscalars~\cite{Chiang:2004nm,Chiang:2006ih}), and that the weak 
phase $\gamma$ is given~\cite{Charles:2006yw,Bona:2006ah}, this provides
information on the magnitude and CP-violating phase of $A_{NP}$ and partial 
information on $\Delta B_{NP}$.

The pair of decays, $B^+\to K^0\pi^+$ and $B^0\to K^+\pi^-$, provides three 
observables, $A_I$ and $A^{+.0}_{CP}$,  which are insufficient 
for determining the four parameters $|\Delta B|, \phi_B, |A'|$ and $\phi_{A'}$.

A recent study~\cite{Gronau:2006ha}, comparing different values measured for 
$A^+_{CP}(B^+\to K^+\pi^0)$ and  $A^0_{CP} (B^0\to K^+\pi^-)$, two processes 
unrelated by isospin-reflection, concluded that a sizable negative strong 
phase difference between $C$ and $T$ amplitudes of comparable magnitude is
required to account for the asymmetries within the SM.  
(In Ref.~\cite{Hou:2005hd,Baek:2007yy} the different asymmetries have been 
argued to provide a clue for NP.) A similar conclusion about ${\rm Arg}(C/T)$
was reached in Ref.~\cite{Buras:2003dj,Chiang:2004nm,Chiang:2006ih} studying
$B\to K\pi$ and $B\to \pi\pi$ decays in flavor SU(3). This seems to stand in
contrast to QCD calculations using a factorization theorem, which obtain small
values for this strong phase 
difference~\cite{Beneke:2001ev,Beneke:2003zv,Bauer:2004tj,Beneke:2005vv}.
In our arguments above we have assumed that ${\rm Arg}(C/T)$ is negligible, in
accordance with QCD calculations. The nonzero strong phase difference measured
in $A^+_{CP}(B^+\to K^+\pi^0)/A^0_{CP}(B^0\to K^+\pi^-)$ may be due to
rescattering from $T$ to $C$ (if $C$ were smaller than $T$ without
rescattering), or the effect of an annihilation amplitude which we included in
$T$.  In either case, the nonzero and negative value of ${\rm Arg}(C/T)$ must
be included in the extraction of NP parameters.
\bigskip

\leftline{\bf C.  $B_s \to K\bar K$} 
\bigskip

The structure of the amplitudes $\Delta B$ and $A$ in the pair 
$(B_s\to K^+K^-, B_s\to K^0\bar K^0)$ is similar to Eqs.~(\ref{DB-Kpi}) 
and (\ref{A-Kpi}), however the term $C+P_{EW}$ is absent 
in $A$~\cite{Gronau:1994rj}. As noted, these processes (or 
$B_s\to K^{*+}K^{*-},$ $B_s\to K^{*0} \bar K^{*0}$) provide in addition to 
an isospin-dependent asymmetry and two direct CP asymmetries also 
the two time-dependent asymmetries $S_{K^+K^-}$ and $S_{K_SK_S}$ 
(or $S_{K^{*+}K^{*-}}$ and $S_{K^{*0}\bar K^{*0}}$) given in 
(\ref{Sk+k-}) and (\ref{Sk0k0}). 
While $S_{K_SK_S}$ is difficult to measure, the first four asymmetries suffice 
for determining the magnitude and weak phase of a potential NP $\Delta I=1$ 
amplitude  and the imaginary part of the $\Delta I=0$ amplitude. A theoretical 
advantage of the pair $(B_s\to K^+K^-, B_s\to K_SK_S)$ over the pair
$(B^+\to K^+\pi^0, B^0\to K^0\pi^0)$ is the absence of SM color-suppressed tree 
($C$) and color-favored EWP ($P_{EW}$) amplitudes in the first pair of 
processes. Thus, in these decays NP amplitudes cannot 
masquerade as unusual EWP amplitudes, as may happen in the second pair of
processes \cite{Gronau:2003kj,Beneke:2003zv,Yoshikawa:2003hb,Buras:2003dj}.
\bigskip

\centerline{\bf VI.  ASPECTS OF SOME SPECIFIC MODELS}
\bigskip

In models with extra isosinglet quarks, such as considered in Refs.\
\cite{Buchmuller:1988et,Grossman:1999av,Bjorken:2002vt,%
Aguilar-Saavedra:2002kr,Andre:2003wc}, mixing of the
isosinglet quark with ordinary quarks can induce flavor-changing neutral
currents (FCNC).  Thus, for example, effective four-fermion operators of the
flavor structure $b \to s q \bar q$ ($q = u,d,s,c$) can arise, with
arbitrary couplings $U_{sb}$ at the $bsZ$ vertex and couplings to $u,d,s,c$
as in the standard electroweak theory.  These differ for left-handed and
right-handed quarks, being given by the interaction Lagrangian
\beq
{\cal L}_{Z \bar f f}=- i\sqrt{g^2+g'^2}\bar f \gamma^\mu (I_{3L}-Q_{\rm EM}
x)Z_\mu f~~~.
\eeq
Here $g$ and $g'$ are the SU(2) and U(1) couplings of the SU(2) $\times$
U(1) electroweak theory, $I_{3L}$ is the left-handed isospin 
and $Q_{\rm EM}$ the charge of fermion $f$, and $x \equiv \sin^2 \theta_W$ 
is the electroweak SU(2)--U(1) mixing parameter.  At values of $Q \le 1$ GeV, 
relevant for color-favored couplings of the neutral current to a light meson,
one expects $x \simeq
0.238$ \cite{Czarnecki:2000ic}, running to $x = 0.23152 \pm 0.00014$ at $Q =
M_Z$ \cite{Yao:2006px}.  (Here we are taking the ``effective'' $x$ as measured
in leptonic asymmetries at $M_Z$.)  The low-energy expectation has been
confirmed by a recent measurement of $x = 0.2397 \pm 0.0010 \pm 0.0008$
at $Q^2 = 0.026$ GeV$^2$ \cite{Anthony:2005pm}.

\renewcommand{\arraystretch}{1.35}
\begin{table}[h]
\caption{Values of charges $I_{3L} - Q_{\rm EM} x,~Q_\chi$, and $Q_\psi$,
governing $Z^{(\prime)} \bar f f$ couplings.  Normalizations are arbitrary
and independent for $Z$, $Z_\chi$, and $Z_\psi$ couplings.
\label{tab:Zchg}}
\begin{center}
\begin{tabular}{r c c c c c c} \hline \hline
    & $u$ & $d$ & $u+d$ & $u-d$ & $e$ & $\nu$ \\ \hline
$Z:~L$ & $\frac12 -\frac23 x$ & $-\frac12 +\frac13 x$ & 
 $-\frac13x$ & $1-x$ & $-\frac12 + x$ & $\frac12$ \\
$R$ & $-\frac23 x$ & $\frac13 x$ & $-\frac13x$ & $-x$ & $x$ & 0\\
$R+L$ & $\frac12-\frac43x$ & $-\frac12+\frac23x$ & $-\frac23x$ &
 $1-2x$ & $-\frac12+2x$ & $\frac12$ \\
$R-L$ & $-\frac12$ & $\frac12$ & 0 & $-1$ & $\frac12$ & $-\frac12$ \\ \hline 
$Z_\chi:~L$ & $-1$ & $-1$ & $-2$ & 0 & 3 & 3 \\
$R$ &  1   & $-3$ & $-2$ & 4 & 1 & 5 \\
$R+L$ & 0 & $-4$ & $-4$ & 4 & 4 & 8 \\ 
$R-L$ & 2 & $-2$ &  0   & 4 & $-2$ & 2 \\ \hline
$Z_\psi:~L$ &  1   &  1   &  2   & 0 & 1 & 1 \\
$R$ & $-1$ & $-1$ & $-2$ & 0 & $-1$ & $-1$ \\
$R+L$ & 0 & 0 & 0 & 0 & 0 & 0 \\
$R-L$ & $-2$ & $-2$ & $-4$ & 0 & $-2$ & $-2$ \\ \hline \hline
\end{tabular}
\end{center}
\end{table}

Many extended grand unified theories can have gauge bosons beyond those of the
SM at relatively low (TeV) masses.  The implications of such bosons for FCNC
processes contributing to $B$ decays have been examined, for example, in
Refs.\ \cite{Barger:2003hg}.  These authors have considered the effects of
operators of various chiralities.  As in the case of FCNC couplings of the $Z$,
a variety of couplings of the $Z'$ to the $q \bar q$ pair in $b \to s q \bar q$
is possible.  The $Z_\chi$, associated with the U(1) in the symmetry-breaking
chain SO(10) $\to$ SU(5) $\times$ U(1)$_\chi$, couples differently to
different SU(5) representations.  As the SU(5) assignments of up-type and
down-type quarks are different, such a boson will have both isoscalar and
isovector couplings.  On the other hand, the $Z_\psi$, associated with the
U(1) in the chain E$_{\rm 6} \to$ SO(10) $\times$ U(1)$_\psi$, couples in
the same way to up- and down-type quarks, and so will have a purely
$\Delta I = 0$ nature.  Couplings of mixtures of $Z_\chi$ and $Z_\psi$, such
as arise in certain versions of superstring compactifications, were discussed
in Ref.\ \cite{London:1986dk}.

The relative $Z_\chi$ charges $Q_\chi$ of SU(5) representations in a
(left-handed) $16^*$-plet of SO(10) are $(3,-1,-5)$ for the $5^* = (\bar d,
e^-,\nu_e)$, $10 = (u,d,\bar u, e^+)$, and $1 = \bar N_e$, respectively.  Here
$\bar N_e$ denotes the (presumably heavy) left-handed antineutrino.  To
calculate the $Z_\chi$ charges of right-handed quarks and leptons, we bear in 
mind that CP-conjugation reverses the sign of all charges.

A simpler rule applies to the $Z_\psi$ charges $Q_\psi$.  In the 27-plet 
of E$_{\rm 6}$ these charges are $(1,-2,4)$ for the 16-, 10-, and 
1-dimensional SO(10) representations.  All the ordinary left-handed 
fermions belong to the 16 of SO(10).  Thus all left-handed $u,d,e,\nu$ 
have charge 1 while right-handed $u,d,e,N_e$ have charge $-1$.

\begin{table}
\caption{Relative couplings $g_M$ and their squares for transitions between
$Z$, $Z_\chi$, or $Z_\psi$ and pseudoscalar or vector mesons.  Normalizations
for $Z$, $Z_\chi$, and $Z_\psi$ are arbitrary and independent of one another;
no relation is implied between pseudoscalar and vector mesons.  For $Z$
couplings to vector mesons we have taken $x=0.238$.
\label{tab:Zcoupl}}
\begin{center}
\begin{tabular}{|r|c c|c c|c c|} \hline \hline
 & \multicolumn{2}{c|}{$Z$} & \multicolumn{2}{c|}{$Z_\chi$} &
   \multicolumn{2}{c|}{$Z_\psi$} \\ \hline
 & $g_M$ & $g_M^2$ & $g_M$ & $g_M^2$ & $g_M$ & $g_M^2$ \\ \hline
$M = \pi^0$ & $\frac{1}{\s}$ & $\frac12$ & $-\frac{4}{\s}$ & 8 & 0 & 0 \\
$\eta$ & $\frac{1}{2\st}$ & $\frac{1}{12}$ & $-\frac{2}{\st}$ & $\frac43$
 & $\frac{2}{\st}$ & $\frac43$ \\
$\eta'$ & $\frac{1}{\sx}$ & $\frac16$ & $-\frac{4}{\sx}$ & $\frac83$
 & $-\frac{8}{\sx}$ & $\frac{32}{3}$ \\ \hline
$M = \rho^0$ & $\frac{1}{\s}\left( 2x-1 \right)$ & 0.137
 & $-\frac{4}{\s}$ & 8 & 0 & 0 \\
$\omega$ & $-\frac{\s}{3}x$ & 0.0126 & $-\frac{4}{\s}$ & 8 & 0 & 0 \\
$\phi$ & $\frac23 x - \frac12$ & 0.117 & $-4$ & 16 & 0 & 0 \\ \hline \hline 
\end{tabular}
\end{center}
\end{table}

In models with FCNC $b \to s$ transitions, whether due to standard $Z$ or $Z'$
exchange, the $Z^{(\prime)} \bar f f$ couplings may be decomposed into $\Delta
I=0$ and $\Delta I=1$ contributions for each chirality (left or right) of
fermion.  Now, matrix elements will depend on fermion chiralities, so the
interpretation of specific ratios of $\Delta I$ amplitudes will depend on
calculations of these matrix elements as in Refs.\ \cite{Grossman:1999av,%
Hiller:2002ci,Atwood:2003tg,Deshpande:2003nx}.  In the last three references,
NP operators are expressed in terms of SM operators ${\cal O}_{3,7,9}$ and
evolved down to low $Q^2$, where additional operators arise.  
The isospin content of flavor-changing $Z$ couplings in $B\to K\pi$ is studied
in Ref.~\cite{Atwood:2003tg}. 

The isospins and chiralities of the $Z \bar f f$ couplings are directly
relevant in the case of contributions to the color-favored electroweak
penguin amplitudes.  In this case the $Z$ couples directly to a meson,
such as $\pi^0,~\eta,~\eta'~,\rho^0,~\omega,~\phi$.  Neglecting QCD corrections
to factorization, we may calculate the relative couplings of a $Z$, $Z_\chi$,
or $Z_\psi$ to various pseudoscalar or various vector mesons using the values
of $I_{3L} - Q_{\rm EM} x,~Q_\chi$, and $Q_\psi$ shown in Table \ref{tab:Zchg}.

The isoscalar ($u+d$) couplings of $Z$ and $Z_\chi$ are purely
vector ($L=R$) while that of $Z_\psi$ is purely axial $(L=-R)$.  The
isovector couplings of $Z$ and $Z_\chi$ contain both vector and axial-vector
contributions, while those of $Z_\psi$ vanish identically.
A pseudoscalar meson couples to
the axial current, while a vector meson couples to the vector current.
We use the quark content $\pi^0 = (d \bar d - u \bar u)/\sqrt{2}$,
$\eta \simeq (s \bar s - u \bar u - d \bar d)/\sqrt{3}$, $\eta' \simeq
(2 s \bar s + u \bar u + d \bar d)/\sqrt{6}$, $\rho^0 = (d \bar d - u \bar u)/
\sqrt{2}$, $\omega = (d \bar d + u \bar u)/\sqrt{2}$, $\phi = s \bar s$.
The results are shown in Table \ref{tab:Zcoupl}.

The relative couplings in Table \ref{tab:Zcoupl} have some interesting
properties.  The standard $Z$ couples more strongly to $\pi^0$ than to $\eta$
or $\eta'$, and much more strongly to $\rho^0$ and $\phi$ than to $\omega$. The
$Z_\chi$ also couples more strongly to $\pi^0$ than to $\eta$ or $\eta'$, but
its coupling to $\phi$ is somewhat stronger than to $\rho^0$ or $\omega$.  The
$Z_\psi$ couples to $\eta$ and overwhelmingly to $\eta'$ but not to $\pi^0$.
By virtue of its purely axial coupling to ordinary matter (members of the
SO(10) 16-plet), it does not couple at all to vector mesons.  As mentioned, QCD
corrections could modify this conclusion.  In fact, QCD corrections have been
calculated for models involving flavor-changing 
Z exchange and were shown to change the above pattern~\cite{Buchalla:2005us}.

FCNC operators which emulate SM electroweak penguins are expected to
contribute also to $B \to X_s \ell^+ \ell^-$ and $B \to X_s \nu \bar \nu$
processes.  The fact that no anomalous behavior of these processes has yet been
seen provides constraints on such operators. The current agreement between SM 
calculations of $\b(B\to X_s\ell^+\ell^-)$~\cite{Ali:2002jg} and experiment
\cite{HFAG,Aubert:2004it}, $\b(B\to X_s\ell^+\ell^-)= 
(4.50^{+1.03}_{-1.01})\times 10^{-6}$ for $M_{\ell^+\ell^-}>0.2~{\rm GeV}/c^2$, 
is somewhat above the level of $1\times 10^{-6}$. For a NP contribution of no
more than this magnitude, this leads to upper bounds on flavor-changing
couplings, such as $|U_{sb}/V_{cb}| < 6\times 10^{-3}$ for the $bsZ$ coupling
in models with an extra isosinglet quark~\cite{Gronau:1996rv}. 
A coupling at this upper limit would contribute less than $1\%$ of the measured 
$B_s$--$\bar
B_s$ mixing~\cite{Abazov:2006dm}, while its EWP-like contributions to $B\to
K\phi$ and $B\to K\pi^0$ would be at a level of several percent of the dominant
penguin amplitude~\cite{Gronau:1996rv}.
 
\newpage
\centerline{\bf VII.  CONCLUSIONS}
\bigskip

We have studied the question of how to determine the isospin structure of
potential NP operators occurring in the effective Hamiltonian describing $b\to
sq\bar q$.  We have shown that this question may be answered by studying four
asymmetries in pairs of isospin-reflected decay processes: an isospin
asymmetry, direct CP asymmetries in $B^+$ and $B^0$ decays, which can be
translated into $\Delta I=0$ and $\Delta I=1$ asymmetries, and a deviation
$\Delta S$ from $\pm\sin 2\beta$ of the coefficient $S$ of the $\sin\Delta mt$
term in time-dependent CP asymmetry. These four observables permit, in
principle, determining the magnitude and CP-violating phase of a $\Delta I=1$
NP amplitude and the imaginary part of a corresponding $\Delta I=0$ amplitude.
Similar considerations apply to $B_s$ decays, where one may use five
asymmetries instead of four for certain pairs of isospin-reflected decays.

The current precision in asymmetry measurements is insufficient for carrying
out this program at this time.  So far, a single nonzero direct CP asymmetry,
$A_{CP}(B^0\to K^+\pi^-) =-0.097\pm 0.012$, has been clearly observed.  The
isospin-reflected pair of processes, $B^+\to K^0\pi^+$ and $B^0\to K^+\pi^-$, 
where $A^{I=0}_{CP}=-0.044\pm 0.014, A^{I=1}_{CP}=0.053 \pm 0.014$
(see Table \ref{tab:asym}), may soon provide first direct evidence for separate
nonzero $\Delta I=0$ and $\Delta I=1$ asymmetries. A sum rule among the four 
$B\to K\pi$ CP asymmetries~\cite{Atwood:1997iw,Gronau:2005kz}, valid
within first order $\Delta I=1$ corrections, predicts~\cite{Gronau:2006xu}
$A_{CP}(B^0\to K^0\pi^0) = -0.140 \pm 0.043$, using the values of the other
three measured asymmetries. This implies a sizable $\Delta I=1$ asymmetry, 
$A^{\Delta I=1}_{CP}=0.094\pm 0.025$, and $A^{\Delta I=0}_{CP}=-0.047\pm 0.025$
in the pair $B^+\to K^+\pi^0, B^0\to K^0\pi^0$. Measurements of the
corresponding isospin asymmetry, $A_I = 0.087\pm 0.038$, and $\Delta S = -0.35
\pm 0.21$ must be improved for a useful implementation of the proposed method.

Certain models of New Physics, such as those involving flavor-changing neutral
currents mediated either by the standard $Z$ or by new neutral gauge
bosons, have distinctive isospin patterns in the operators giving rise to
deviations from Standard Model observables.  We have given examples of some
of these patterns for models with additional singlet quarks or with the
gauge bosons $Z_\chi,Z_\psi$ of SO(10) and E$_{\rm 6}$ theories.   

\bigskip

\centerline{\bf ACKNOWLEDGMENTS}
\bigskip

J. L. R. is grateful to the Lewiner Institute for Theoretical Physics for its
support, and wishes to thank the Physics Department of the Technion for their
gracious hospitality during part of this investigation.  We thank Tim Gershon
and Jim Smith for information on updated data.  This work was
supported in part by the United States Department of Energy under Grant
No.\ DE FG02 90ER40560, by the Israel Science Foundation
under Grant No.\ 1052/04, and by the German-Israeli Foundation under
Grant No.\ I-781-55.14/2003.


\begin{thebibliography}{99}

\bibitem{Gronau:1996rv}
  M.~Gronau and D.~London,
  Phys.\ Rev.\ D {\bf 55}, 2845 (1997)
  [arXiv:hep-ph/9608430].

\bibitem{Grossman:1996ke}
  Y.~Grossman and M.~P.~Worah,
  Phys.\ Lett.\ B {\bf 395}, 241 (1997)
  [arXiv:hep-ph/9612269].
  
\bibitem{Ciuchini:1997zp} 
M.~Ciuchini, E.~Franco, G.~Martinelli, A.~Masiero and L.~Silvestrini,
Phys.\ Rev.\ Lett.\  {\bf 79}, 978 (1997)
[arXiv:hep-ph/9704274].
  
\bibitem{London:1997zk}
  D.~London and A.~Soni,
  Phys.\ Lett.\ B {\bf 407}, 61 (1997)
 [arXiv:hep-ph/9704277].

\bibitem{Barbieri:1997kq}
  R.~Barbieri and A.~Strumia,
  Nucl.\ Phys.\ B {\bf 508}, 3 (1997)
  [arXiv:hep-ph/9704402].

\bibitem{Fleischer:2001pc}
  R.~Fleischer and T.~Mannel,
  Phys.\ Lett.\ B {\bf 511}, 240 (2001)
  [arXiv:hep-ph/0103121].
 See also R.~Fleischer and T.~Mannel,
  Phys.\ Lett.\ B {\bf 506}, 311 (2001)
  [arXiv:hep-ph/0101276].

\bibitem{Gronau:1998ep}
  M.~Gronau and J.~L.~Rosner,
  Phys.\ Rev.\ D {\bf 59}, 113002 (1999)
  [arXiv:hep-ph/9809384].
 
\bibitem{Lipkin:1998ie}
  H.~J.~Lipkin,
  Phys.\ Lett.\ B {\bf 445}, 403 (1999)
  [arXiv:hep-ph/9810351].
  
\bibitem{Atwood:1997iw}
  D.~Atwood and A.~Soni,
  Phys.\ Rev.\ D {\bf 58}, 036005 (1998)
  [arXiv:hep-ph/9712287].
  
  \bibitem{Matias:2001ch}
  J.~Matias,
  Phys.\ Lett.\  B {\bf 520}, 131 (2001)
  [arXiv:hep-ph/0105103].

\bibitem{Gronau:2005gz} M.~Gronau and J.~L.~Rosner,
  Phys.\ Rev.\  D {\bf 71}, 074019 (2005) [arXiv:hep-ph/0503131].

\bibitem{Gronau:2005kz} M.~Gronau,
  Phys.\ Lett.\ B {\bf 627}, 82 (2005) [arXiv:hep-ph/0508047].

 \bibitem{Gronau:2006eb}
  M.~Gronau, Y.~Grossman, G.~Raz and J.~L.~Rosner,
  Phys.\ Lett.\ B {\bf 635}, 207 (2006)
  [arXiv:hep-ph/0601129].

\bibitem{London:1989ph}
  D.~London and R.~D.~Peccei,
  Phys.\ Lett.\ B {\bf 223}, 257 (1989).
  
 \bibitem{Cheng:2005bg} These deviations have been estimated, for instance,  
 in H.~Y.~Cheng, C.~K.~Chua and A.~Soni,
  Phys.\ Rev.\  D {\bf 72}, 014006 (2005) [arXiv:hep-ph/0502235];
  M.~Beneke,
  Phys.\ Lett.\  B {\bf 620}, 143 (2005) [arXiv:hep-ph/0505075];
  M.~Gronau, J.~L.~Rosner and J.~Zupan,
  Phys.\ Rev.\  D {\bf 74}, 093003 (2006) [arXiv:hep-ph/0608085].

  \bibitem{HFAG} Heavy Flavor Averaging Group, E. Barberio 
{\it et al.}, hep-ex/0603003, regularly updated in 
{\tt http://www.slac.stanford.edu/xorg/hfag/}.

\bibitem{Lipkin:1991st}
H. J. Lipkin, Y. Nir, H. R. Quinn, and A. Snyder, Phys.\ Rev.\ D {\bf 44},
1454 (1991).

\bibitem{Gronau:1991dq}
  M.~Gronau,
  Phys.\ Lett.\ B {\bf 265}, 389 (1991).

\bibitem{Gronau:2003ep}
  M.~Gronau and J.~L.~Rosner,
  Phys.\ Lett.\ B {\bf 564}, 90 (2003)
  [arXiv:hep-ph/0304178].

\bibitem{Aubert:2007ub}
  B.~Aubert  {\it et al.} [BABAR Collaboration],
  arXiv:hep-ex/0702010.

\bibitem{Gronau:2005ax}
  M.~Gronau and J.~L.~Rosner,
  Phys.\ Rev.\ D {\bf 72}, 094031 (2005)
  [arXiv:hep-ph/0509155].
  
\bibitem{Gronau:1994rj}
  M.~Gronau, O.~F.~Hernandez, D.~London and J.~L.~Rosner,
  Phys.\ Rev.\  D {\bf 50}, 4529 (1994)
  [arXiv:hep-ph/9404283];
  Phys.\ Rev.\  D {\bf 52}, 6374 (1995)
  [arXiv:hep-ph/9504327].
  
\bibitem{Ciuchini:1997rj}
  M.~Ciuchini, R.~Contino, E.~Franco, G.~Martinelli and L.~Silvestrini,
  Nucl.\ Phys.\  B {\bf 512}, 3 (1998)
  [Erratum-ibid.\  B {\bf 531}, 656 (1998)]
  [arXiv:hep-ph/9708222].
  
\bibitem{Keum:2000wi}
  Y.~Y.~Keum, H.~N.~Li and A.~I.~Sanda,
  Phys.\ Rev.\  D {\bf 63}, 054008 (2001)
  [arXiv:hep-ph/0004173].
  
\bibitem{Beneke:2001ev}
  M.~Beneke, G.~Buchalla, M.~Neubert and C.~T.~Sachrajda,
  Nucl.\ Phys.\  B {\bf 606}, 245 (2001)
  [arXiv:hep-ph/0104110].
  
\bibitem{Bauer:2005kd}
  C.~W.~Bauer, I.~Z.~Rothstein and I.~W.~Stewart,
   Phys.\ Rev.\ D {\bf 74}, 034010 (2006)
  [arXiv:hep-ph/0510241].
  
\bibitem{Aubert:2007mb}
  B.~Aubert  {\it et al.} [BaBar Collaboration],
  arXiv:hep-ex/0702043.
  
\bibitem{Aubert:2007me}
  B.~Aubert {\it et al.} [BABAR Collaboration],
  arXiv:hep-ex/0702046.
  
\bibitem{Garmash:2004wa}
  A.~Garmash {\it et al.}  [BELLE Collaboration],
  Phys.\ Rev.\ D {\bf 71}, 092003 (2005)
  [arXiv:hep-ex/0412066].

\bibitem{Aubert:2006nu}
  B.~Aubert {\it et al.}  [BABAR Collaboration],
  Phys.\ Rev.\ D {\bf 74}, 032003 (2006)
  [arXiv:hep-ex/0605003].
  B.~Aubert {\it et al.}  [BABAR Collaboration],
  arXiv:hep-ex/0607112.
  
\bibitem{Abe:2005qe} K.~Abe {\it et al.}  [BELLE Collaboration],
  arXiv:hep-ex/0508052.
  
\bibitem{Aubert:2006qu}  B.~Aubert {\it et al.}  [BABAR Collaboration],
  Phys.\ Rev.\ D {\bf 74}, 011106 (2006)  [arXiv:hep-ex/0603040].

\bibitem{Aubert:2005ce}  B.~Aubert {\it et al.}  [BABAR Collaboration],
  Phys.\ Rev.\ D {\bf 72}, 072003 (2005)
  [Erratum-ibid.\ D {\bf 74}, 099903 (2006)]  [arXiv:hep-ex/0507004];
  Phys.\ Rev.\ D {\bf 73}, 031101 (2006)
  [arXiv:hep-ex/0508013];
 A.~Garmash {\it et al.},
  Phys.\ Rev.\ Lett.\  {\bf 96}, 251803 (2006)
  [arXiv:hep-ex/0512066].
  
\bibitem{Gronau:2003kj}
M.~Gronau and J.~L.~Rosner,
  Phys.\ Lett.\ B {\bf 572}, 43 (2003)
  [arXiv:hep-ph/0307095].
  
\bibitem{Beneke:2003zv}
  M.~Beneke and M.~Neubert,
  Nucl.\ Phys.\ B {\bf 675}, 333 (2003)
  [arXiv:hep-ph/0308039].

\bibitem{Gronau:2006xu}
  M.~Gronau and J.~L.~Rosner,
  Phys.\ Rev.\ D {\bf 74}, 057503 (2006)
  [arXiv:hep-ph/0608040].
  
  \bibitem{Gronau:1989ia}
  M.~Gronau,
  Phys.\ Rev.\ Lett.\  {\bf 63}, 1451 (1989).
  
\bibitem{Aleksan:1994if}
  R.~Aleksan, B.~Kayser and D.~London,
  Phys.\ Rev.\ Lett.\  {\bf 73}, 18 (1994)
  [arXiv:hep-ph/9403341].
  
\bibitem{Silva:1996ih}
  J.~P.~Silva and L.~Wolfenstein,
  Phys.\ Rev.\  D {\bf 55}, 5331 (1997)
  [arXiv:hep-ph/9610208].

\bibitem{Datta:2004re}
  A.~Datta and D.~London,
  Phys.\ Lett.\  B {\bf 595}, 453 (2004)
  [arXiv:hep-ph/0404130].

\bibitem{Baek:2004rp}
  S.~Baek, P.~Hamel, D.~London, A.~Datta and D.~A.~Suprun,
  Phys.\ Rev.\  D {\bf 71}, 057502 (2005)
  [arXiv:hep-ph/0412086].

\bibitem{Datta:2004jm}
  A.~Datta, M.~Imbeault, D.~London, V.~Page, N.~Sinha and R.~Sinha,
  Phys.\ Rev.\  D {\bf 71}, 096002 (2005)
  [arXiv:hep-ph/0406192].
  
\bibitem{Dighe:1997wj} A.~S.~Dighe, M.~Gronau and J.~L.~Rosner,
  Phys.\ Rev.\  D {\bf 57}, 1783 (1998) [arXiv:hep-ph/9709223].
  
\bibitem{Chen:2001pr} C.~H.~Chen, Y.~Y.~Keum and H.~n.~Li,
  Phys.\ Rev.\  D {\bf 64} (2001) 112002 [arXiv:hep-ph/0107165].
  
\bibitem{Chiang:2003pm}
  C.~W.~Chiang, M.~Gronau, Z.~Luo, J.~L.~Rosner and D.~A.~Suprun,
  Phys.\ Rev.\  D {\bf 69}, 034001 (2004)
  [arXiv:hep-ph/0307395].
  
\bibitem{Chiang:2004nm}
  C.~W.~Chiang, M.~Gronau, J.~L.~Rosner and D.~A.~Suprun,
  Phys.\ Rev.\  D {\bf 70}, 034020 (2004)
  [arXiv:hep-ph/0404073].

\bibitem{Chiang:2006ih} C.~W.~Chiang and Y.~F.~Zhou,
  JHEP {\bf 0612}, 027 (2006) [arXiv:hep-ph/0609128].

\bibitem{Charles:2006yw}
J.~Charles {\it et al.} [CKMfitter Collaboration], eConf {\bf C060409}, 043 
(2006), presenting updated results periodically on the web site 
{\tt http://www.slac.stanford.edu/xorg/ckmfitter/}.

\bibitem{Bona:2006ah}
  M.~Bona {\it et al.}  [UTfit Collaboration],
  JHEP {\bf 0610}, 081 (2006)
  [arXiv:hep-ph/0606167],
  presenting updated results periodically on the web site 
{\tt http://www.utfit.org/}.
 
\bibitem{Gronau:2006ha} M.~Gronau and J.~L.~Rosner,
  Phys.\ Lett.\  B {\bf 644}, 237 (2007) [arXiv:hep-ph/0610227].

\bibitem{Hou:2005hd} W.~S.~Hou, M.~Nagashima and A.~Soddu,
  Phys.\ Rev.\ Lett.\  {\bf 95}, 141601 (2005)
  [arXiv:hep-ph/0503072].

\bibitem{Baek:2007yy} S.~Baek and D.~London,
  arXiv:hep-ph/0701181.
  
\bibitem{Bauer:2004tj}
  C.~W.~Bauer, D.~Pirjol, I.~Z.~Rothstein and I.~W.~Stewart,
  Phys.\ Rev.\  D {\bf 70}, 054015 (2004)
  [arXiv:hep-ph/0401188];
C.~W.~Bauer, I.~Z.~Rothstein and I.~W.~Stewart,
 Phys.\ Rev.\ Lett.\  {\bf 94}, 231802 (2005) [arXiv:hep-ph/0412120].

\bibitem{Beneke:2005vv} M.~Beneke and S.~Jager,
  Nucl.\ Phys.\  B {\bf 751}, 160 (2006) [arXiv:hep-ph/0512351].
  
 \bibitem{Yoshikawa:2003hb}
  T.~Yoshikawa,
  Phys.\ Rev.\  D {\bf 68}, 054023 (2003)
  [arXiv:hep-ph/0306147].
  
  \bibitem{Buras:2003dj}
  A.~J.~Buras, R.~Fleischer, S.~Recksiegel and F.~Schwab,
  Phys.\ Rev.\ Lett.\  {\bf 92}, 101804 (2004)
  [arXiv:hep-ph/0312259];
  Nucl.\ Phys.\ B {\bf 697}, 133 (2004)
  [arXiv:hep-ph/0402112].
  
\bibitem{Buchmuller:1988et}
  W.~Buchmuller and M.~Gronau,
  Phys.\ Lett.\  B {\bf 220}, 641 (1989).
  
\bibitem{Grossman:1999av} Y.~Grossman, M.~Neubert and A.~L.~Kagan,
  JHEP {\bf 9910}, 029 (1999) [arXiv:hep-ph/9909297].

\bibitem{Bjorken:2002vt} J.~D.~Bjorken, S.~Pakvasa and S.~F.~Tuan,
  Phys.\ Rev.\  D {\bf 66}, 053008 (2002) [arXiv:hep-ph/0206116].

\bibitem{Aguilar-Saavedra:2002kr} J.~A.~Aguilar-Saavedra,
  Phys.\ Rev.\  D {\bf 67}, 035003 (2003)
  [Erratum-ibid.\  D {\bf 69}, 099901 (2004)] [arXiv:hep-ph/0210112].

\bibitem{Andre:2003wc} T.~C.~Andre and J.~L.~Rosner,
  Phys.\ Rev.\  D {\bf 69}, 035009 (2004) [arXiv:hep-ph/0309254].

\bibitem{Czarnecki:2000ic} A.~Czarnecki and W.~J.~Marciano,
  Int.\ J.\ Mod.\ Phys.\ A {\bf 15}, 2365 (2000)
  [arXiv:hep-ph/0003049].

\bibitem{Yao:2006px} W.~M.~Yao {\it et al.} [Particle Data Group],
  J.\ Phys.\ G {\bf 33}, 1 (2006).

\bibitem{Anthony:2005pm} P.~L.~Anthony {\it et al.} [SLAC E158 Collaboration],
  Phys.\ Rev.\ Lett.\  {\bf 95}, 081601 (2005)
  [arXiv:hep-ex/0504049].

\bibitem{Barger:2003hg} V.~Barger, C.~W.~Chiang, P.~Langacker and H.~S.~Lee,
Phys.\ Lett.\  B {\bf 580}, 186 (2004) [arXiv:hep-ph/0310073];
{\it ibid.} {\bf 598}, 218 (2004):
V.~Barger, C.~W.~Chiang, J.~Jiang and P.~Langacker,
Phys.\ Lett.\  B {\bf 596}, 229 (2004) [arXiv:hep-ph/0405108].

\bibitem{London:1986dk} D.~London and J.~L.~Rosner,
Phys.\ Rev.\ D {\bf 34}, 1530 (1986).

\bibitem{Hiller:2002ci} G.~Hiller,
  Phys.\ Rev.\  D {\bf 66}, 071502 (2002)
  [arXiv:hep-ph/0207356].

\bibitem{Atwood:2003tg} D.~Atwood and G.~Hiller,
  arXiv:hep-ph/0307251.

\bibitem{Deshpande:2003nx} N.~G.~Deshpande and D.~K.~Ghosh,
  Phys.\ Lett.\  B {\bf 593}, 135 (2004)
  [arXiv:hep-ph/0311332].
  
\bibitem{Buchalla:2005us}
  G.~Buchalla, G.~Hiller, Y.~Nir and G.~Raz,
  JHEP {\bf 0509}, 074 (2005)
  [arXiv:hep-ph/0503151].
 
\bibitem{Ali:2002jg} A.~Ali, E.~Lunghi, C.~Greub and G.~Hiller,
   Phys.\ Rev.\  D {\bf 66}, 034002 (2002) [arXiv:hep-ph/0112300];
  A.~Ghinculov, T.~Hurth, G.~Isidori and Y.~P.~Yao,
  Nucl.\ Phys.\  B {\bf 685}, 351 (2004) [arXiv:hep-ph/0312128].

\bibitem{Aubert:2004it} B.~Aubert {\it et al.} [BABAR Collaboration],
  Phys.\ Rev.\ Lett.\  {\bf 93}, 081802 (2004) [arXiv:hep-ex/0404006];
  M.~Iwasaki {\it et al.} [Belle Collaboration],
  Phys.\ Rev.\  D {\bf 72}, 092005 (2005) [arXiv:hep-ex/0503044].
  
\bibitem{Abazov:2006dm}
  V.~M.~Abazov {\it et al.}  [D0 Collaboration],
  Phys.\ Rev.\ Lett.\  {\bf 97}, 021802 (2006)
  [arXiv:hep-ex/0603029];
  A.~Abulencia {\it et al.}  [CDF Collaboration],
  Phys.\ Rev.\ Lett.\  {\bf 97}, 242003 (2006)
  [arXiv:hep-ex/0609040].
  
\end{thebibliography}
\end{document}